\documentclass[sigconf,nonacm]{acmart}
\AtBeginDocument{%
  }

\copyrightyear{2026}
\acmYear{2026}
\setcopyright{cc}
\setcctype{by}
\usepackage{amsmath}
\usepackage{algorithmic}
\usepackage{booktabs}
\usepackage{multirow}
\usepackage{booktabs}
\usepackage{graphicx}
\usepackage{threeparttable}
\usepackage{graphicx}  % for \resizebox
\usepackage{graphicx}
\usepackage{url}
\usepackage{hyperref}
\usepackage{placeins}
\usepackage{textcomp}
\usepackage{orcidlink}
\usepackage{xcolor}
\usepackage{tcolorbox}
\tcbuselibrary{skins, breakable}
\usepackage{enumitem}
\usepackage{amsmath}
\usepackage{tcolorbox}
\tcbuselibrary{skins, breakable}
\usepackage[T1]{fontenc}
\usepackage[utf8]{inputenc}
\usepackage{xcolor}
\usepackage{tcolorbox}

\definecolor{RoyalBlueBox}{RGB}{0,51,153}
\begin{document}

%\title{Automated Methods for Rapid Bug Triage in the TianoCore Open-Source Software Ecosystem}
\title{TianoForge: An Automated Bug Triage Approach for the TianoCore UEFI Firmware Development Community}
\author{Nazanin Siavash}
\email{nsiavash@uccs.edu}
\orcid{0009-0000-4177-0632}
\affiliation{
  \institution{University of Colorado Colorado Springs (UCCS)}
  %\city{Colorado Spring}
  \state{Colorado}
  \country{United States}
}

\author{Terrance E. Boult}
\email{tboult@uccs.edu}
\orcid{0000-0001-5007-2529}
\affiliation{
  \institution{University of Colorado Colorado Springs (UCCS)}
  %\city{Colorado Spring}
  \state{Colorado}
  \country{United States}
}

\author{Armin Moin}
\email{moin@purdue.edu}
\orcid{0000-0002-8484-7836}
\affiliation{
  \institution{Purdue University}
  %\city{Colorado Spring}
  \state{Indiana}
  \country{United States}
}

\renewcommand{\shortauthors}{Siavash et al.}

\begin{abstract}
  We propose a novel approach to bug triage in the TianoCore open-source UEFI firmware development ecosystem. This integrated approach, called \textit{TianoForge}, deploys the state of the art in artificial intelligence, specifically machine learning, to enable automated bug triage. This includes invalid bug report detection, duplicate bug report detection, bug report prioritization, and bug report assignment. We use various Generative Pretrained Transformer (GPT) Large Language Models (LLMs) with and without Retrieval Augmented Generation (RAG) to automate these tasks. Given the crucial role of bug triage in software maintenance and the huge number of untriaged issues in the TianoCore community, in particular, their primary project, EDK II, we expect a significant impact on the efficiency of TianoCore software maintenance processes, primarily bug triage and resolution. Our experimental study shows that TianoForge reduces the average bug triage time from around 11 days to approximately 7 minutes, which is a 99.95\% reduction.
\end{abstract}

%Bug triage is a critical yet time-consuming activity in software maintenance, involving tasks such as invalid issue detection, duplicate detection, priority prediction, and developer assignment. Existing research typically addresses these tasks in isolation, while integrated automated triage solutions remain largely unexplored, particularly in firmware ecosystems such as TianoCore. 

\begin{CCSXML}
<ccs2012>
 <concept>
  <concept_id>10011007.10011074.10011111.10011113</concept_id>
  <concept_desc>Software and its engineering~Software maintenance tools</concept_desc>
  <concept_significance>500</concept_significance>
 </concept>
 <concept>
  <concept_id>10002978.10003022.10003465</concept_id>
  <concept_desc>Security and privacy~Software and application security</concept_desc>
  <concept_significance>300</concept_significance>
 </concept>
</ccs2012>
\end{CCSXML}

\ccsdesc[500]{Software and its engineering~Software maintenance tools}
\ccsdesc[300]{Security and privacy~Software and application security}
\keywords{bug triage, uefi, firmware, tianocore, edk ii, large language models}


\maketitle

\section{Introduction}\label{Introduction}
Modern software systems evolve continuously to accommodate changing requirements and performance improvements, inevitably introducing defects \cite{HeYang2021, SerranoCiordia2005}. To manage these defects, bug tracking systems (e.g., Bugzilla) are widely used to collect and organize reports submitted by users and developers. These reports provide essential information for reproducing and resolving issues; however, efficiently processing them remains a significant challenge.
Bug triage, the process of evaluating reported issues and assigning them to suitable developers for resolution \cite{Anvik+2006}, is a critical yet resource-intensive task. In large-scale and open-source projects, the high volume of incoming reports makes manual triage impractical \cite{AnvikMurphy2011}. Delays in assignment, coupled with frequent reassignment (i.e., bug tossing), substantially increase resolution time and reduce overall efficiency. Moreover, accurate triage requires detailed knowledge of developer expertise and system components, which is difficult to maintain in dynamic, distributed teams \cite{DaoYang2023}. As a result, manual triage often suffers from low accuracy and scalability limitations \cite{Wang+2024}.
Automated bug triage has therefore become essential in modern software engineering \cite{Chen+2019,Su+2021}. By leveraging data-driven and intelligent techniques, automated approaches can improve triage efficiency through report classification, prioritization, deduplication, and assignment \cite{Bocu+2023,Liu+2023}. These methods reduce manual effort, shorten resolution cycles, and improve diagnostic precision by filtering noisy or redundant reports. Additionally, structured triage outputs enable downstream automation tasks, such as root cause analysis and failure prediction, supporting more proactive system maintenance.
Beyond efficiency, the quality and trustworthiness of bug reports themselves present an emerging challenge. Recent discussions within the TianoCore EDK~II community highlight concerns about undisclosed 
LLM-assisted bug reports, where AI-generated content introduced technical inaccuracies and undermined maintainer trust in issue-tracking 
data~\cite{edk2_issue11747}. This underscores the importance of 
domain-aware automated triage tools that can assist developers in 
identifying and filtering low-quality or invalid reports, rather than relying solely on contributor-submitted content at face value.

Despite significant advances in automated bug triage, existing approaches primarily focus on one or two triage tasks in isolation, such as bug assignment, duplicate detection, or prioritization, rather than providing an integrated end-to-end triage solution. Moreover, most prior studies have been conducted on general-purpose software projects, while the TianoCore EDK II ecosystem has not previously been studied in the context of automated bug triage. As a result, the effectiveness of recent Large Language Model (LLM)-based approaches in integrated bug triage pipelines remains largely unexplored, particularly in the context of firmware development. Furthermore, the impact of retrieval augmentation, prompt engineering, and domain-specific knowledge on LLM-assisted triage performance is not yet well understood.

In this work, we propose \textit{\textbf{TianoForge}}, an integrated automated bug triage script for the TianoCore EDK II ecosystem. TianoForge streamlines the bug triage process by automatically performing four key triage tasks: invalid detection, duplicate detection, priority prediction, and developer assignment within a unified workflow. We conduct an experimental study to address the following Research Questions (RQs):
\textbf{RQ1:} Can automated bug triage reduce the bug resolution time compared to manual triage in TianoCore Projects, specifically EDK~II?
\textbf{RQ2:} How effective and efficient is the proposed automated triage framework across key sub-tasks (e.g., duplicate detection, prioritization, invalid report detection, and assignment)?
\textbf{RQ3:} What are the key factors in deciding the priority level of the issues in TianoCore projects?
\textbf{RQ4:} To what extent can prompt engineering improve the performance of LLM-based triage components?
%\textbf{RQ4:} Can deploying Large Language Models (LLMs) in the automated triage pipeline increase the effectiveness and efficiency of the triage process? %RQ4 & RQ2 were the same

The remainder of this paper is organized as follows. Section~\ref{Background} provides background on bug triage, LLMs, and Retrieval-Augmented Generation (RAG). Section~\ref{related-work} reviews existing research on bug triage tasks. Section~\ref{proposed-approach} presents TianoForge, including the problem formulation, architecture, and integrated triage workflow. Section~\ref{experimental-study} describes the experimental dataset, evaluation metrics, experimental setup, and results analysis. Section~\ref{discussion} discusses the main findings and implications of the results. Finally, Section~\ref{conclusion-future} concludes the paper and outlines directions for future work.

\section{Background}\label{Background}
This section introduces the key concepts underlying the proposed automated bug triage framework. It first discusses bug triage and bug report enhancement, highlighting their role in managing software defects and improving issue quality. It then reviews LLMs and prompt engineering techniques that support automated reasoning over bug reports. Finally, it presents RAG, which enables LLMs to leverage external project-specific knowledge during inference.
\subsection{Bug Triage and Report Enhancement}

Bug tracking systems, such as Bugzilla, Jira, and GitHub Issues, are essential tools for managing software defects throughout the development lifecycle. These systems allow users and developers to report, track, and resolve issues in a structured manner. A typical bug report consists of a textual summary (title), a detailed description, and additional metadata such as priority, severity, component, and assigned developer. Over time, reports may also include comments, attachments, and status updates that reflect the debugging process.

The large volume of reports generated in modern software projects makes efficient issue management increasingly challenging. As a result, bug triage has become a critical process for ensuring that defects are analyzed and addressed effectively. Bug triage involves evaluating reported issues and determining appropriate actions, including invalid issue detection, duplicate detection, issue prioritization, and developer assignment \cite{Lee+2022}. These tasks help development teams allocate resources efficiently and reduce the time required to resolve reported issues.

The effectiveness of bug triage is highly dependent on the quality of the underlying bug reports. Incomplete descriptions, ambiguous titles, missing metadata, and redundant reports can hinder decision-making and reduce triage accuracy. To address these challenges, bug report enhancement techniques are often employed to improve report quality prior to or during the triage process. Such techniques may include refining issue titles, enriching reports with additional context, and filtering invalid or duplicate submissions. By improving the quality of input reports, enhancement techniques can support more accurate and consistent triage decisions.

\subsection{Large Language Models and Prompt Engineering}

The growing complexity and scale of modern software repositories have motivated the development of automated approaches for bug triage. Recent advances in LLMs have made them promising candidates for supporting such automation. LLMs are transformer-based foundation models trained on massive text corpora to perform a wide range of natural language understanding and generation tasks \cite{SiavashMoin2025}. Their ability to capture semantic relationships in unstructured text makes them well suited for analyzing bug reports and assisting with triage-related decisions.

In software engineering, LLMs have been applied to tasks such as code generation, software maintenance, defect prediction, and bug triage \cite{PhillipsGloria2025}. A key factor influencing their effectiveness is prompt engineering, which refers to the design of instructions and contextual information provided to the model. Techniques such as zero-shot and few-shot prompting allow LLMs to perform specialized tasks without additional model training. Furthermore, prompts can incorporate domain knowledge, decision criteria, and representative examples to guide model behavior and improve prediction quality \cite{Fan+2023}.

\subsection{Retrieval-Augmented Generation}

While LLMs possess strong reasoning and language understanding capabilities, they are limited by the knowledge encoded in their training data. RAG addresses this limitation by incorporating external information during inference. Instead of relying solely on model parameters, RAG retrieves relevant documents from an external knowledge source and includes them as additional context in the prompt.

In bug triage applications, retrieval sources may include historical bug reports, project documentation, developer discussions, and previously resolved issues. By grounding predictions in project-specific information, RAG can provide additional context for tasks such as duplicate detection, prioritization, and developer assignment. Consequently, the combination of LLMs, prompt engineering, and retrieval augmentation has emerged as a promising direction for developing more effective automated bug triage systems.

\section{Related Work}\label{related-work}
Bug triage is a fundamental task in software maintenance and closely interacts with related activities such as duplicate detection, prioritization, and bug validity assessment. Early research primarily formulated bug triage as a supervised text classification or optimization problem over bug report content and metadata \cite{Ahsan+2009}. Traditional approaches employed techniques such as Naïve Bayes classifiers, topic models, and bug tossing graphs to model developer expertise and report characteristics \cite{Xuan+2010,BhattacharyaNeamtiu2010,Xia+2016}.
In recent years, research has shifted toward deep learning–based methods, representation learning, and hybrid architectures that leverage richer semantic and structural information from bug reports.
Graph-based feature augmentation has been explored to enhance textual representations of bug reports. For instance, Alazzam et al. constructed term graphs where words are treated as nodes and enriched via neighborhood relationships, significantly improving bug prioritization performance compared to traditional TF-IDF and baseline deep learning methods \cite{Alazzam+2020}. Similarly, Umer et al. \cite{Umer+2019} proposed a CNN-based prioritization framework that integrates syntactic, semantic, and emotional features, demonstrating substantial improvements in cross-project F1-scores and highlighting the importance of domain-specific signals in triage-related tasks.
Hierarchical models have also been introduced to better capture the structure of bug reports. Yadav and Rathore \cite{YadavRathore2024} proposed a Hierarchical Attention Network (HAN) that models both word-level and sentence-level representations using DistilBERT tokenization and Bi-GRU layers. Their approach outperforms classical machine learning models, including SVM, random forest, and logistic regression, as well as standard deep learning baselines, demonstrating the effectiveness of hierarchical attention mechanisms.
Beyond individual tasks, recent work has explored integrated pipelines for bug management. Chhabra and Chadha \cite{ChhabraChadha2025} introduced ReBaRF-Bug, which combines TF-IDF features, multi-stage data augmentation (e.g., SMOTE and paraphrasing), ResNet-based feature transformation, and ensemble learning. Their approach achieved high accuracy and robustness on datasets such as Eclipse and Mozilla, indicating that combining augmentation with deep feature extraction can effectively address data imbalance and sparsity.
In specialized domains, such as security bug classification, cross-project learning, and similarity-based augmentation have been widely adopted. Ganzorig et al. \cite{Ganzorig+2025} leveraged both lexical and embedding-based similarity methods to retrieve related bug reports from external projects and train deep learning models, including CNNs, LSTMs, GRUs, and Transformers. Their results demonstrated consistently high performance, although no single model dominated across all scenarios.
More recently, Large Language Models (LLMs) have emerged as a promising paradigm for automated bug triage. Compared to traditional representations such as TF-IDF and static embeddings, transformer-based models provide richer contextual representations, leading to improved performance in tasks such as bug assignment and prioritization. These models are particularly effective when leveraging both bug titles and detailed descriptions. Dipongkor \cite{Dipongkor2024} investigated transformer-based LLMs for bug triaging and developer assignment, evaluating several models including BERT, RoBERTa, CodeBERT, and DeBERTa. Their study demonstrated that transformer-based models outperform traditional approaches and further showed that ensemble voting strategies can improve assignment accuracy beyond individual models. More recently, Kiashemshaki et al. \cite{Kiashemshaki+2025} proposed an instruction-tuned LLM framework for automated bug triaging using a LoRA-adapted DeepSeek-R1 model with candidate-constrained decoding. Their approach generates ranked developer recommendations directly from bug reports without requiring handcrafted features or graph construction, demonstrating the feasibility of lightweight project-specific LLMs for practical bug assignment tasks. You et al. \cite{You+2025} proposed an ``LLM-as-classifier'' paradigm, introducing semi-supervised and human-in-the-loop strategies for hierarchical text classification, with applications including proactive bug triage and monitoring systems.

Hybrid architectures that combine LLMs with structural learning methods have also gained attention. Song et al. \cite{Song+2025} proposed a model that integrates LLM-derived representations with multi-scale feature fusion and graph neural networks (GNNs). By capturing both global semantic context and local structural relationships, their approach achieves consistent improvements across multiple evaluation metrics, suggesting strong potential for complex triage scenarios where relational information is important.
\section{Proposed Approach}\label{proposed-approach}
This section presents \textit{TianoForge}, a unified LLM-based script for automated bug triage in TianoCore EDK~II. TianoForge integrates invalid issue detection, duplicate detection, bug prioritization, and developer assignment within a single workflow.

\subsection{Problem Formulation}
Let $\mathcal{G} = \{b_1, b_2, \ldots, b_N\}$ ($N = 75$) denote the set of
GitHub-native bugs submitted to the EDK~II repository, where each
bug $b_k$ is represented by the feature tuple
\[
  b_k = \bigl(\mathit{title}_k,\; \mathit{package}_k,\; \mathit{type}_k,\;
         \mathit{priority}_k,\; \mathit{comments}_k,\; \mathit{state}_k\bigr).
\]
Let $\mathcal{G}^L \subseteq \mathcal{G}$ ($|\mathcal{G}^L| = 37$) denote
the subset of bugs with a known first assignee. Let $\mathcal{B}$ denote
the historical Bugzilla XML corpus used as a retrieval source
in System~B. Each task is solved by prompting an LLM in an in-context
learning setting under two systems: \textbf{System~A} and
\textbf{System~B}. The four tasks are
described below in pipeline order.

\textbf{Retrieval notation.}
All retrieval-augmented systems use a hybrid ranker combining a dense
semantic retriever and a sparse lexical retriever. Let
$\mathrm{Dense}(b_k, \mathcal{X})$ denote the ranked list of bugs from
corpus $\mathcal{X}$ ordered by cosine similarity to the BGE-base-en-v1.5
embedding of $b_k$, and let $\mathrm{BM25}(b_k, \mathcal{X})$ denote the
ranked list ordered by BM25 lexical score. Reciprocal Rank Fusion (RRF)
combines both lists:
\[
  \mathrm{RRF}(A, B)_i \;=\; \frac{1}{k_0 + \mathrm{rank}_A(i)}
                             + \frac{1}{k_0 + \mathrm{rank}_B(i)},
  \quad k_0 = 60,
\]
where $\mathrm{rank}_A(i)$ is the rank of item $i$ in list $A$.
The notation $\mathrm{RRF}(\cdot,\cdot)_{1:\kappa}$ denotes the
top-$\kappa$ items by RRF score.
\subsubsection{Invalid Bug Detection}
Invalid bug detection aims to identify bug reports that do not correspond
to genuine software defects.

\textbf{Definition.}
Each bug $b_k \in \mathcal{G}$ is associated with a binary label
$y^{\mathrm{inv}}_k \in \{0,1\}$. The objective is to predict:
\[
  f_{\mathrm{inv}} : \mathcal{G} \;\rightarrow\; \{0,\;1\}, \qquad
  \hat{y}^{\mathrm{inv}}_k = f_{\mathrm{inv}}(b_k),
\]
where $\hat{y}^{\mathrm{inv}}_k = 1$ indicates the bug is \emph{invalid}
and $\hat{y}^{\mathrm{inv}}_k = 0$ indicates the bug is \emph{valid}.
\begin{tcolorbox}[
  enhanced, breakable,
  colback=RoyalBlueBox!5,
  colframe=RoyalBlueBox,
  colbacktitle=RoyalBlueBox,
  coltitle=white,
  fonttitle=\bfseries\small,
  title={Invalid Bug Criteria},
  left=4pt, right=4pt, top=3pt, bottom=3pt,
  boxrule=0.8pt, arc=4pt
]

A bug $b_k$ is labeled invalid ($\hat{y}^{\mathrm{inv}}_k = 1$) if it
satisfies one or more of the following:
\begin{enumerate}[nosep, leftmargin=*, label=(\roman*)]
  \item spam or entirely unrelated to firmware development;
  \item a general usage question rather than a bug report or feature request;
  \item describes behavior working as designed;
  \item lacks sufficient information to reproduce or investigate.
\end{enumerate}
When in doubt, the LLM classifies the bug as \emph{valid}.
\end{tcolorbox}
\textbf{LLM Predictor.}
The function $f_{\mathrm{inv}}$ receives
$(\mathit{title}_k,\allowbreak\
\mathit{package}_k,\allowbreak\
\mathit{type}_k,\allowbreak\
\mathit{comments}_k,\allowbreak\
\mathit{state}_k)$
with the \texttt{invalid} state label masked to prevent leakage. \textbf{System~A} provides $\kappa = 3$
fixed few-shot examples per class drawn from $\mathcal{B}$, identical
across all queries. \textbf{System~B} retrieves the top-$\kappa = 3$ most
similar bugs from $\mathcal{B}$ per query via RRF:
\[
  \mathcal{E}_k = \mathrm{RRF}\bigl(
    \mathrm{Dense}(b_k,\mathcal{B}),\;
    \mathrm{BM25}(b_k,\mathcal{B})
  \bigr)_{1:\kappa}, \quad \kappa = 3.
\]
\subsubsection{Duplicate Bug Detection}
Duplicate bug detection determines whether a newly submitted bug describes an issue already reported in the repository.

\textbf{Definition.}
Given a query bug $b_k \in \mathcal{G}$ and candidate bugs
$\mathcal{C}_k \subset \mathcal{G} \backslash \{b_k\}$, an LLM
$f_{\mathrm{dup}}$ predicts:
\[
  \hat{y}^{\mathrm{dup}}_k \;=\; f_{\mathrm{dup}}(b_k,\;\mathcal{C}_k)
  \;\in\; \{0,\;1\},
\]
where $\hat{y}^{\mathrm{dup}}_k = 1$ indicates $b_k$ is a duplicate of
some $b_j \in \mathcal{C}_k$.

\textbf{Candidate retrieval.}
Candidates are obtained via RRF over BGE-base-en-v1.5 and BM25, both
indexed over $\mathcal{G}$:
\[
  \mathcal{C}_k = \mathrm{RRF}\bigl(
    \mathrm{Dense}(b_k,\mathcal{G}),\;
    \mathrm{BM25}(b_k,\mathcal{G})
  \bigr)_{1:\kappa}, \quad \kappa = 5.
\]

  \begin{tcolorbox}[
  enhanced, breakable,
  colback=RoyalBlueBox!5,
  colframe=RoyalBlueBox,
  colbacktitle=RoyalBlueBox,
  coltitle=white,
  fonttitle=\bfseries\small,
  title={Duplicate Bug Criterion},
  left=4pt, right=4pt, top=3pt, bottom=3pt,
  boxrule=0.8pt, arc=4pt
]

Two bugs $b_k$ and $b_j$ are duplicates if and only if they describe the
\textbf{same root cause} in the same codebase component. Topical similarity
or a shared package is insufficient. When in doubt, the LLM returns
$\hat{y}^{\mathrm{dup}}_k = 0$.
\end{tcolorbox}

\textbf{LLM Predictor.}
Both systems retrieve candidate bugs $\mathcal{C}_k$ via RRF as described
above. \textbf{System~A} prompts the LLM with $b_k$ and $\mathcal{C}_k$
only. \textbf{System~B} additionally retrieves up to $\kappa' = 3$
confirmed duplicate pairs from $\mathcal{B}$ via RRF as in-context
demonstrations.
\subsubsection{Bug Prioritization}

Bug priority classification automatically determines the urgency of a bug
report, helping developers allocate resources efficiently.

\textbf{Definition.}
Let $\mathcal{P} = \{\texttt{low}, \texttt{medium}, \texttt{high}\}$ be
the set of priority levels. The task is formulated as a multi-class
classification problem where an LLM $f_{\mathrm{pri}}$ predicts:
\[
  f_{\mathrm{pri}} : \mathcal{G} \;\rightarrow\; \mathcal{P}, \qquad
  \hat{y}^{\mathrm{pri}}_k \;=\;
  \arg\max_{p_j \in \mathcal{P}}\; P(p_j \mid b_k)
\]

\begin{tcolorbox}[
  enhanced, breakable,
  colback=RoyalBlueBox!5,
  colframe=RoyalBlueBox,
  colbacktitle=RoyalBlueBox,
  coltitle=white,
  fonttitle=\bfseries\small,
  title={Priority Level Definitions},
  left=4pt, right=4pt, top=3pt, bottom=3pt,
  boxrule=0.8pt, arc=4pt
]
\begin{description}[nosep, leftmargin=2em, style=nextline]
  \item[\texttt{high}] Crashes, data corruption, security vulnerabilities
        within the UEFI threat model, or build failures blocking \emph{all}
        configurations for all users.
  \item[\texttt{medium}] Functional bugs with workarounds, compiler-specific
        build failures, or missing features required for spec compliance.
  \item[\texttt{low}] Cosmetic defects, code-quality improvements, warnings
        not blocking builds, optional features, or documentation changes.
\end{description}
\end{tcolorbox}
\textbf{LLM Predictor}
The function $f_{\mathrm{pri}}$ is prompted with $b_k$'s features and
domain-calibration rules encoding EDK~II-specific priority heuristics
(e.g., CVE patches are not automatically \texttt{high} under the UEFI
threat model; compiler-specific build failures are \texttt{medium}).
\textbf{System~A} uses Leave-One-Out (LOO): all $N - 1$ bugs in
$\mathcal{G}$ with known priority labels serve as in-context examples.
\textbf{System~B} uses the same LOO examples as System~A and additionally
retrieves the top-$\kappa$ Bugzilla bugs with known priorities via RRF:
\[
  \mathcal{R}_k = \mathrm{RRF}\bigl(
    \mathrm{Dense}(b_k,\mathcal{B}^{\mathcal{P}}),\;
    \mathrm{BM25}(b_k,\mathcal{B}^{\mathcal{P}})
  \bigr)_{1:\kappa}, \quad \kappa = 5,
\]
where $\mathcal{B}^{\mathcal{P}} \subseteq \mathcal{B}$ is the Bugzilla
corpus filtered to bugs with $\mathit{priority} \in \mathcal{P}$.

%\textbf{Output.} $\hat{y}^{\mathrm{pri}}_k \in \mathcal{P}$.
\subsubsection{Bug Assignment}
Bug assignment aims to automatically recommend/predict the most appropriate developer for resolving a newly submitted bug.

\textbf{Definition.}
Let $\mathcal{D} = \{d_1, d_2, \ldots, d_{|\mathcal{D}|}\}$ be the closed
set of developers observed in $\mathcal{G}^L$. The task is formulated as a
multi-class classification problem where an LLM $f_{\mathrm{asgn}}$
predicts:
\[
  f_{\mathrm{asgn}} : \mathcal{G}^L \;\rightarrow\; \mathcal{D}, \qquad
  \hat{y}^{\mathrm{asgn}}_k \;=\;
  \arg\max_{d_j \in \mathcal{D}}\; P(d_j \mid b_k).
\]
\begin{tcolorbox}[
  enhanced, breakable,
  colback=RoyalBlueBox!5,
  colframe=RoyalBlueBox,
  colbacktitle=RoyalBlueBox,
  coltitle=white,
  fonttitle=\bfseries\small,
  title={Assignment Constraint},
  left=4pt, right=4pt, top=3pt, bottom=3pt,
  boxrule=0.8pt, arc=4pt
]

The predicted assignee must belong to the closed set $\mathcal{D}$.
Predictions outside $\mathcal{D}$ are corrected by case-insensitive exact
matching; if no match is found, the prediction defaults to the first
assignee in $\mathcal{D}$. The LLM is provided with package-level
assignment heuristics encoding domain knowledge of the edk2 codebase
(e.g., \texttt{CryptoPkg} CLANG build errors $\to$ \texttt{mdkinney};
\texttt{NetworkPkg} CodeQL bugs $\to$ \texttt{BritChesley};
\texttt{OvmfPkg} physical memory bugs $\to$ \texttt{os-d}).
\end{tcolorbox}
\textbf{LLM Predictor}
The function $f_{\mathrm{asgn}}$ is prompted with $b_k$'s features,
package-level heuristics, and in-context labelled examples.
\textbf{System~A} uses LOO: the prompt contains all $|\mathcal{G}^L| - 1$
labelled examples from $\mathcal{G}^L$ excluding $b_k$.
\textbf{System~B} retrieves the top-$\kappa$ most similar bugs from
$\mathcal{G}^L$ via RRF:
\[
  \mathcal{S}_k = \mathrm{RRF}\bigl(
    \mathrm{Dense}(b_k,\mathcal{G}^L),\;
    \mathrm{BM25}(b_k,\mathcal{G}^L)
  \bigr)_{1:\kappa}, \quad \kappa = 5.
\]
\subsubsection{Integrated Triage Script}
The four tasks above are composed into a single sequential script operating on $\mathcal{G}^L$, enabling end-to-end triage from submission to developer assignment.

\textbf{Design}
The framework shares a single ChromaDB vector store and BM25 index across
all four tasks, eliminating redundant indexing. A unified \texttt{call\_llm}
function routes requests to the OpenAI Chat Completions API, OpenAI
Responses API (GPT-5+), or Anthropic Messages API based on the model name.
The best-performing model for each task is active by default:
\textbf{System~A} with \texttt{claude-\allowbreak sonnet-\allowbreak 4-6} for Tasks~1--3 and
\texttt{gpt-5.5} for Task~4.

\textbf{Output.}
For each $b_k \in \mathcal{G}^L$:
$\hat{y}^{\mathrm{inv}}_k \in \{0,1\}$,\;
$\hat{y}^{\mathrm{dup}}_k \in \{0,1\}$,\;
$\hat{y}^{\mathrm{pri}}_k \in \mathcal{P}$,\;
$\hat{y}^{\mathrm{asgn}}_k \in \mathcal{D}$.
%\subsubsection{Scoring Metrics}
%\cite{AcharyaGinde2024}
%\cite{Mashhadi+2023}
%\cite{MarjaiKiss2025}
\subsection{Architecture of the Proposed Solution}
\begin{figure*}[t]
\centering
\includegraphics[width=0.6\textwidth]{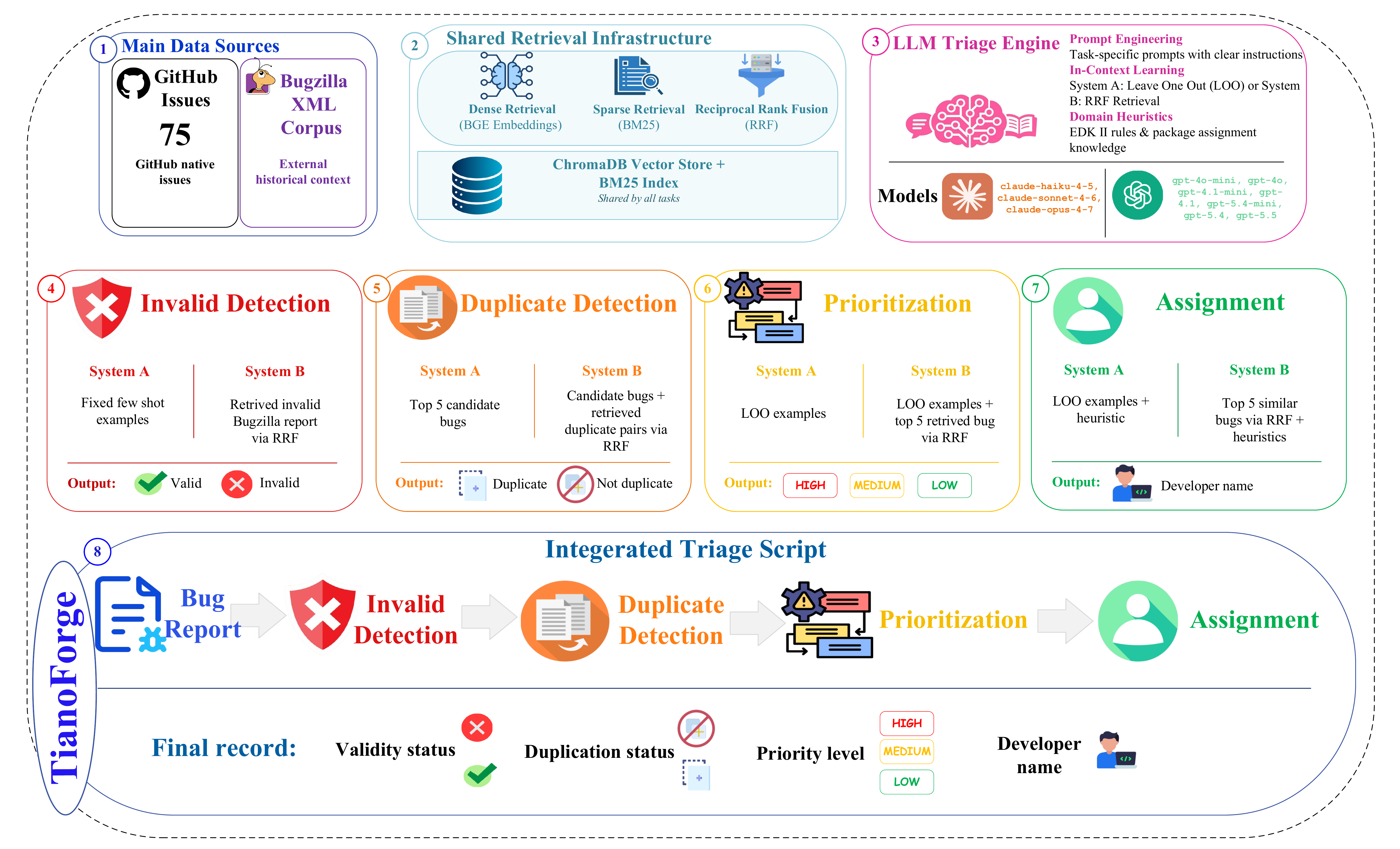}\caption{Overview of TianoForge}
\label{fig:overview}
\end{figure*}

Figure~\ref{fig:overview} presents the architecture of \textbf{TianoForge}, an integrated automated bug triage script for the TianoCore EDK~II ecosystem. TianoForge combines four key triage tasks, invalid detection, duplicate detection, priority prediction, and developer assignment, into a unified sequential workflow that automatically processes submitted bug reports and produces a complete triage record.
The architecture consists of three main components: (1) data sources, (2) a shared retrieval and reasoning layer, and (3) task-specific triage modules. The primary input is a GitHub-native bug report, while a historical Bugzilla XML corpus serves as an external knowledge source. The GitHub dataset contains the issues being triaged, whereas the Bugzilla corpus provides historical context used for retrieval-augmented prompting RAG configurations.

To support retrieval-augmented triage, TianoForge employs a shared retrieval infrastructure consisting of a dense semantic retriever based on BGE-base-en-v1.5 embeddings, a sparse lexical retriever based on BM25, and Reciprocal Rank Fusion (RRF) for combining retrieval results. Both retrieval mechanisms operate over a shared ChromaDB vector store and BM25 index constructed from the historical Bugzilla corpus. This shared infrastructure eliminates redundant indexing and enables all triage tasks to access relevant historical information through a common retrieval layer.

The retrieved context is consumed by a unified LLM triage engine responsible for executing all four triage tasks. The engine incorporates task-specific prompt engineering, in-context learning, and domain heuristics derived from the EDK~II ecosystem. Depending on the task and experimental configuration, prompts are executed using models from the OpenAI and Anthropic families.

The first stage of the script performs \textit{invalid detection}. Given a bug report, the model determines whether the issue represents a valid software defect or should be considered invalid. System~A relies on fixed few-shot examples, whereas System~B augments the prompt with retrieved invalid reports from the Bugzilla corpus.

The second stage performs \textit{duplicate detection}. Candidate duplicate reports are first retrieved using the hybrid retrieval layer. System~A uses only the retrieved candidate bugs, while System~B additionally incorporates retrieved duplicate pairs from the Bugzilla XML corpus as contextual examples. The output is a binary decision indicating whether the issue is a duplicate of an existing report.

The third stage performs \textit{prioritization}. This component classifies issues into low, medium, or high priority levels using domain-calibrated EDK~II priority criteria. System~A utilizes LOO examples, while System~B supplements these examples with top-ranked historical bugs retrieved from the Bugzilla corpus. The resulting priority label reflects the urgency.
The final stage performs \textit{developer assignment}. Using package-level assignment heuristics and historical examples, the model predicts the most suitable developer to resolve the issue. System~A relies on LOO examples and assignment heuristics, whereas System~B retrieves the top-$\kappa$ most similar bugs from the 
labeled GitHub-native subset $\mathcal{G}^L$ via RRF as in-context 
demonstrations

The four tasks are executed sequentially within the \textbf{TianoForge} script. Starting from a bug report, the script produces a final triage record consisting of the validity status, duplicate status, priority level, and recommended developer. By integrating these traditionally independent triage activities into a single automated workflow, TianoForge enables rapid and consistent bug triage while leveraging both project-specific domain knowledge and historical repository information.

\section{Experimental Study}\label{experimental-study}
\subsection{Experimental Dataset}
The EDK II dataset is constructed from the publicly available issue tracking
system of the TianoCore EDK II project hosted on GitHub. To the best of our
knowledge, this dataset has not been previously utilized in the context of
this research problem. The dataset is collected and curated on
\textbf{April 3, 2026}, ensuring that the repository state is fixed at that
point in time.

Data collection is performed programmatically using the GitHub REST API
through a custom Python script, retrieving issues in a paginated manner with
the following constraints: (i) only issues (excluding pull requests) are
retrieved, (ii) only closed issues are included, and (iii) only issues
labelled as \texttt{type:bug} are selected.

The final dataset consists of 2,610 issues structured as a CSV
file with the following fields: issue number, title, state, first assignee,
first assignment timestamp, triage hours, creation and closure timestamps,
URL, milestone, comment count, and three label-derived categorical features:
\texttt{package}, \texttt{priority}, and \texttt{state}.
The dataset comprises two subsets: \textit{2,535 Bugzilla-transferred
issues} (historical bugs migrated from the legacy Bugzilla tracker) and
\textit{75 GitHub-native issues} (bugs filed directly on GitHub). The
GitHub-native issues span issues created between December 2024 and
February 2026. Their priority labels are distributed as 32 medium, 28 low,
and 15 high. Of the 75 GitHub-native issues, 37 carry a known first
assignee and form the labelled subset $\mathcal{G}^L$ used for supervised
evaluation of bug assignment.
\subsection{Evaluation Metrics}

All four tasks are evaluated using standard classification metrics. Let
$y_k$ denote the ground truth label and $\hat{y}_k$ the predicted label
for bug $b_k$. For a given class $c$, let $\mathrm{TP}_c$,
$\mathrm{FP}_c$, $\mathrm{FN}_c$, and $\mathrm{TN}_c$ denote the number
of True Positives, False Positives, False Negatives, and True Negatives,
respectively.

\subsubsection{Accuracy} measures the fraction of correctly classified instances
over all $n$ evaluated bugs:
\[
  \mathrm{Acc} = \frac{1}{n} \sum_{k=1}^{n} \mathbf{1}[\hat{y}_k = y_k]
\]

\subsubsection{Precision} for class $c$ measures the fraction of predicted
positives that are truly positive:
\[
  P_c = \frac{\mathrm{TP}_c}{\mathrm{TP}_c + \mathrm{FP}_c}.
\]

\subsubsection{Recall} for class $c$ measures the fraction of true positives
that are correctly identified:
\[
  R_c = \frac{\mathrm{TP}_c}{\mathrm{TP}_c + \mathrm{FN}_c}.
\]

\subsubsection{F1-Score} for class $c$ is the harmonic mean of precision and
recall:
\[
  F1_c = \frac{2 \cdot P_c \cdot R_c}{P_c + R_c}
\]

For all four tasks, two aggregation strategies are reported.\\
\textbf{Macro} averaging computes the unweighted mean across all classes:
\[
  P_{\mathrm{macro}} = \frac{1}{|\mathcal{C}|} \sum_{c \in \mathcal{C}} P_c,
  \quad
  R_{\mathrm{macro}} = \frac{1}{|\mathcal{C}|} \sum_{c \in \mathcal{C}} R_c,
  \quad
  F1_{\mathrm{macro}} = \frac{1}{|\mathcal{C}|} \sum_{c \in \mathcal{C}} F1_c
\]
\textbf{Weighted} averaging weights each class by its support
$n_c$ (number of true instances of class $c$):
\[
\begin{aligned}
P_{\mathrm{weighted}} &= \frac{1}{n} \sum_{c \in \mathcal{C}} n_c P_c,
\quad
R_{\mathrm{weighted}} &= \frac{1}{n} \sum_{c \in \mathcal{C}} n_c R_c,\\
F1_{\mathrm{weighted}} &= \frac{1}{n} \sum_{c \in \mathcal{C}} n_c F1_c.
\end{aligned}
\]
\subsubsection{Resolution Time} provides a practical efficiency metric
complementing the classification measures above. The total resolution time
of a bug $b_k$ is defined as:
\[
  T^{\mathrm{res}}_k = T^{\mathrm{triage}}_k + T^{\mathrm{fix}}_k,
\]
where $T^{\mathrm{triage}}_k$ is the \emph{triage time} (elapsed time from
issue creation to first assignment) and $T^{\mathrm{fix}}_k$ is the
\emph{fix time} (elapsed time from first assignment to issue closure).
Since $T^{\mathrm{fix}}_k$ depends on developer effort and is unaffected
by the triage pipeline, reducing $T^{\mathrm{triage}}_k$ directly reduces
$T^{\mathrm{res}}_k$.

The mean triage time across the 37 labeled GitHub-native issues
$\mathcal{G}^L$ is $\bar{T}^{\mathrm{triage}} = 260.57$ hours
(10.86 days), reflecting the manual triage latency in the
current process. We compare this against the elapsed time of the
integrated triage script measured using Python's \texttt{time.time()},
which records three intervals: setup time (CSV loading and index
construction), inference time (LLM API calls across all four tasks), and
total elapsed time (setup + inference). Each measurement is averaged over
three independent runs to account for the non-deterministic nature of
LLM inference. TianoForge produces all four predictions for all 37
issues in a single automated run, and any reduction in triage time
directly translates to a reduction in overall resolution time
$T^{\mathrm{res}}_k$.

\subsection{Experimental Setup}
All experiments are conducted on Google Colab. The pipeline is implemented
in Python and relies on the following key libraries: \texttt{openai},
\texttt{anthropic}, \texttt{chromadb}, \texttt{sentence-transformers},
\texttt{rank\_bm25}, \texttt{scikit-learn}, and \texttt{tqdm}.

Ten LLMs are evaluated across all tasks, spanning two providers, OpenAI and Anthropic. For inference, temperature is set to $0.0$ for all OpenAI models to
ensure deterministic outputs. Claude models are called with default
settings as the temperature parameter. 
For retrieval, we use \texttt{BAAI/bge-base-en-v1.5}
as the dense embedding model, indexed via ChromaDB with cosine similarity.
Sparse retrieval uses \texttt{BM25Okapi}. Both are combined via Reciprocal
Rank Fusion with $k_0 = 60$. The top-$\kappa$ values are $\kappa = 5$ for
duplicate candidates, priority context, and assignment context, and
$\kappa' = 3$ for invalid detection context and duplicate pair
demonstrations.
\subsection{Experimental Analysis}
This section presents the experimental results for each triage sub-task and the integrated script. All results are averaged over three independent runs to account for LLM non-determinism. 
\subsubsection{Duplicate Bug Detection}
 
Table~\ref{tab:dup_gt0_avg} reports results under the zero-duplicate ground
truth, where no labeled duplicate pairs exist among the 75 GitHub-native
bugs. Under this assumption, a model that never flags any issue achieves
perfect positive-class precision (P+ = 1.0), and accuracy reflects the
fraction of issues correctly left unflagged. Three models namely
\texttt{gpt-4o-mini} System~A, \texttt{claude-sonnet-4-6} System~A/B, and
\texttt{claude-opus-4-7} System~A flag zero issues consistently across
all three runs, achieving Acc = 0.973 and Mac-F1 = 0.493.
\texttt{claude-sonnet-4-6} is the most robust, maintaining this conservative
behaviour on both System~A and System~B, while most other models produce
false positives on System~B.
 
Table~\ref{tab:dup_pairbased_avg} reports results under the pair-based ground
truth comprising two confirmed duplicate pairs: $\{11791, 11790\}$ and \linebreak
$\{11948, 11462\}$. These pairs are identified through a manual review
process that goes beyond the scope of automated bug triage: we inspect every
issue flagged as a duplicate by the LLMs across all runs and models, and
manually verify whether the flag is correct by reading the issue content and
comparing it against the reported original. Issues confirmed as genuine
duplicates are added to the ground truth. This process serves a dual purpose; it enables a more meaningful evaluation of duplicate detection performance, and it contributes to the \textit{data quality improvement} of the TianoCore EDK~II repository.
 
Evaluation is pair-based, meaning a true positive is scored when either member of a
pair is correctly flagged as a duplicate of the other. All models achieve
R+ = 0.500, meaning every model detects exactly one of the two pairs on
average. The differentiating factor is false positives. On System~A, three
models achieve FP = 0 across all runs --- \texttt{gpt-4o-mini},
\texttt{claude-sonnet-4-6}, and \texttt{claude-opus-4-7} --- yielding
P+ = 1.000, Mac-F1 = 0.826, and Acc = 0.960. On System~B,
\texttt{claude-sonnet-4-6} is the only model achieving FP = 0.
System~A consistently outperforms System~B for most models, as Bugzilla RAG
context generally increases false positives rather than reducing them,
suggesting that historical duplicate pair examples from Bugzilla do not transfer well to the GitHub issue domain.
 
\begin{table*}[ht]
\centering
\caption{Duplicate Detection Results --- Average over 3 Runs (Ground Truth: Zero Duplicates)}
\label{tab:dup_gt0_avg}
\renewcommand{\arraystretch}{0.50} % reduces vertical row spacing
\setlength{\aboverulesep}{1pt}
\setlength{\belowrulesep}{1pt}
\resizebox{\textwidth}{!}{%
\begin{tabular}{llccccccccccccc}
\toprule
\textbf{Model} & \textbf{Sys} & \textbf{P+} & \textbf{R+} & \textbf{F1+} & \textbf{P-} & \textbf{R-} & \textbf{F1-} & \textbf{Mac-P} & \textbf{Mac-R} & \textbf{Mac-F1} & \textbf{Wgt-P} & \textbf{Wgt-R} & \textbf{Wgt-F1} & \textbf{Acc} \\
\midrule
\multirow{2}{*}{gpt-4o-mini}      & A & \textbf{1.000} & 1.000 & \textbf{1.000} & \textbf{1.000} & \textbf{0.973} & \textbf{0.987} & \textbf{1.000} & \textbf{0.987} & \textbf{0.493} & \textbf{1.000} & \textbf{0.973} & \textbf{0.987} & \textbf{0.973} \\
                                   & B & 0.000 & 1.000 & 0.000 & 1.000 & 0.924 & 0.961 & 0.500 & 0.962 & 0.480 & 1.000 & 0.924 & 0.961 & 0.924 \\
\midrule
\multirow{2}{*}{gpt-4o}           & A & 0.000 & 1.000 & 0.000 & 1.000 & 0.938 & 0.968 & 0.500 & 0.969 & 0.484 & 1.000 & 0.938 & 0.968 & 0.938 \\
                                   & B & 0.000 & 1.000 & 0.000 & 1.000 & 0.742 & 0.852 & 0.500 & 0.871 & 0.426 & 1.000 & 0.742 & 0.852 & 0.742 \\
\midrule
\multirow{2}{*}{gpt-4.1-mini}     & A & 0.000 & 1.000 & 0.000 & 1.000 & 0.947 & 0.973 & 0.500 & 0.973 & 0.486 & 1.000 & 0.947 & 0.973 & 0.947 \\
                                   & B & 0.000 & 1.000 & 0.000 & 1.000 & 0.907 & 0.951 & 0.500 & 0.953 & 0.476 & 1.000 & 0.907 & 0.951 & 0.907 \\
\midrule
\multirow{2}{*}{gpt-4.1}          & A & 0.000 & 1.000 & 0.000 & 1.000 & 0.960 & 0.980 & 0.500 & 0.980 & 0.490 & 1.000 & 0.960 & 0.980 & 0.960 \\
                                   & B & 0.000 & 1.000 & 0.000 & 1.000 & 0.893 & 0.944 & 0.500 & 0.947 & 0.472 & 1.000 & 0.893 & 0.944 & 0.893 \\
\midrule
\multirow{2}{*}{gpt-5.4-mini}     & A & 0.000 & 1.000 & 0.000 & 1.000 & 0.884 & 0.939 & 0.500 & 0.942 & 0.469 & 1.000 & 0.884 & 0.939 & 0.884 \\
                                   & B & 0.000 & 1.000 & 0.000 & 1.000 & 0.942 & 0.970 & 0.500 & 0.971 & 0.485 & 1.000 & 0.942 & 0.970 & 0.942 \\
\midrule
\multirow{2}{*}{gpt-5.4}          & A & 0.000 & 1.000 & 0.000 & 1.000 & 0.920 & 0.958 & 0.500 & 0.960 & 0.479 & 1.000 & 0.920 & 0.958 & 0.920 \\
                                   & B & 0.000 & 1.000 & 0.000 & 1.000 & 0.902 & 0.949 & 0.500 & 0.951 & 0.474 & 1.000 & 0.902 & 0.949 & 0.902 \\
\midrule
\multirow{2}{*}{gpt-5.5}          & A & 0.000 & 1.000 & 0.000 & 1.000 & 0.956 & 0.977 & 0.500 & 0.978 & 0.489 & 1.000 & 0.956 & 0.977 & 0.956 \\
                                   & B & 0.000 & 1.000 & 0.000 & 1.000 & 0.924 & 0.961 & 0.500 & 0.962 & 0.480 & 1.000 & 0.924 & 0.961 & 0.924 \\
\midrule
\multirow{2}{*}{claude-haiku-4-5} & A & 0.000 & 1.000 & 0.000 & 1.000 & 0.885 & 0.939 & 0.500 & 0.942 & 0.469 & 1.000 & 0.885 & 0.939 & 0.885 \\
                                   & B & 0.000 & 1.000 & 0.000 & 1.000 & 0.951 & 0.975 & 0.500 & 0.976 & 0.488 & 1.000 & 0.951 & 0.975 & 0.951 \\
\midrule
\multirow{2}{*}{claude-sonnet-4-6}& A & \textbf{1.000} & 1.000 & \textbf{1.000} & \textbf{1.000} & \textbf{0.973} & \textbf{0.987} & \textbf{1.000} & \textbf{0.987} & \textbf{0.493} & \textbf{1.000} & \textbf{0.973} & \textbf{0.987} & \textbf{0.973} \\
                                   & B & \textbf{1.000} & 1.000 & \textbf{1.000} & \textbf{1.000} & \textbf{0.973} & \textbf{0.987} & \textbf{1.000} & \textbf{0.987} & \textbf{0.493} & \textbf{1.000} & \textbf{0.973} & \textbf{0.987} & \textbf{0.973} \\
\midrule
\multirow{2}{*}{claude-opus-4-7}  & A & \textbf{1.000} & 1.000 & \textbf{1.000} & \textbf{1.000} & \textbf{0.973} & \textbf{0.987} & \textbf{1.000} & \textbf{0.987} & \textbf{0.493} & \textbf{1.000} & \textbf{0.973} & \textbf{0.987} & \textbf{0.973} \\
                                   & B & 0.000 & 1.000 & 0.000 & 1.000 & 0.960 & 0.980 & 0.500 & 0.980 & 0.490 & 1.000 & 0.960 & 0.980 & 0.960 \\
\bottomrule
\end{tabular}%
}
%\begin{tablenotes}
%\small
%\item P+/R+/F1+ = Precision, Recall, F1 for the duplicate (positive) class. P-/R-/F1- = for the non-duplicate (negative) class. Mac = macro-average. Wgt = weighted average (support: 0 duplicates, 75 non-duplicates). Under this ground truth, P+ = 1.0 when no issues are flagged, R+ = 1.0 always (FN = 0). Accuracy reflects the fraction of issues correctly not flagged. Bold = models that flagged zero issues across all runs.
%\end{tablenotes}
\end{table*}
\begin{table*}[ht]
\centering
\caption{Duplicate Detection Results --- Average over 3 Runs (Ground Truth: 2 Pairs, \{11791,11790\} and \{11948,11462\})}
\label{tab:dup_pairbased_avg}
\renewcommand{\arraystretch}{0.50} % reduces vertical row spacing
\setlength{\aboverulesep}{1pt}
\setlength{\belowrulesep}{1pt}
\resizebox{\textwidth}{!}{%
\begin{tabular}{llccccccccccccc}
\toprule
\textbf{Model} & \textbf{Sys} & \textbf{P+} & \textbf{R+} & \textbf{F1+} & \textbf{P-} & \textbf{R-} & \textbf{F1-} & \textbf{Mac-P} & \textbf{Mac-R} & \textbf{Mac-F1} & \textbf{Wgt-P} & \textbf{Wgt-R} & \textbf{Wgt-F1} & \textbf{Acc} \\
\midrule
\multirow{2}{*}{gpt-4o-mini}      & A & \textbf{1.000} & 0.500 & \textbf{0.667} & \textbf{0.973} & \textbf{1.000} & \textbf{0.986} & \textbf{0.986} & \textbf{0.750} & \textbf{0.826} & \textbf{0.973} & \textbf{0.986} & \textbf{0.977} & \textbf{0.960} \\
                                   & B & 0.217 & 0.500 & 0.302 & 0.971 & 0.948 & 0.960 & 0.594 & 0.724 & 0.631 & 0.951 & 0.936 & 0.942 & 0.911 \\
\midrule
\multirow{2}{*}{gpt-4o}           & A & 0.278 & 0.500 & 0.356 & 0.972 & 0.962 & 0.967 & 0.625 & 0.731 & 0.661 & 0.953 & 0.950 & 0.950 & 0.924 \\
                                   & B & 0.072 & 0.667 & 0.129 & 0.976 & 0.762 & 0.856 & 0.524 & 0.714 & 0.492 & 0.952 & 0.759 & 0.836 & 0.742 \\
\midrule
\multirow{2}{*}{gpt-4.1-mini}     & A & 0.333 & 0.500 & 0.400 & 0.972 & 0.972 & 0.972 & 0.653 & 0.736 & 0.686 & 0.954 & 0.959 & 0.956 & 0.933 \\
                                   & B & 0.170 & 0.500 & 0.253 & 0.971 & 0.930 & 0.950 & 0.570 & 0.715 & 0.601 & 0.949 & 0.918 & 0.931 & 0.893 \\
\midrule
\multirow{2}{*}{gpt-4.1}          & A & 0.500 & 0.500 & 0.500 & 0.972 & 0.986 & 0.979 & 0.736 & 0.743 & 0.740 & 0.959 & 0.973 & 0.966 & 0.947 \\
                                   & B & 0.143 & 0.500 & 0.222 & 0.970 & 0.916 & 0.942 & 0.557 & 0.708 & 0.582 & 0.948 & 0.904 & 0.922 & 0.880 \\
\midrule
\multirow{2}{*}{gpt-5.4-mini}     & A & 0.137 & 0.500 & 0.211 & 0.970 & 0.902 & 0.934 & 0.553 & 0.701 & 0.573 & 0.947 & 0.891 & 0.915 & 0.871 \\
                                   & B & 0.306 & 0.500 & 0.378 & 0.972 & 0.967 & 0.969 & 0.639 & 0.734 & 0.674 & 0.953 & 0.954 & 0.953 & 0.929 \\
\midrule
\multirow{2}{*}{gpt-5.4}          & A & 0.200 & 0.500 & 0.286 & 0.971 & 0.944 & 0.957 & 0.586 & 0.722 & 0.621 & 0.950 & 0.932 & 0.939 & 0.907 \\
                                   & B & 0.159 & 0.500 & 0.241 & 0.970 & 0.925 & 0.947 & 0.565 & 0.712 & 0.594 & 0.948 & 0.913 & 0.928 & 0.889 \\
\midrule
\multirow{2}{*}{gpt-5.5}          & A & 0.444 & 0.500 & 0.467 & 0.972 & 0.981 & 0.977 & 0.708 & 0.741 & 0.722 & 0.958 & 0.968 & 0.963 & 0.942 \\
                                   & B & 0.217 & 0.500 & 0.302 & 0.971 & 0.948 & 0.960 & 0.594 & 0.724 & 0.631 & 0.951 & 0.936 & 0.942 & 0.911 \\
\midrule
\multirow{2}{*}{claude-haiku-4-5} & A & 0.141 & 0.500 & 0.216 & 0.970 & 0.906 & 0.937 & 0.555 & 0.703 & 0.577 & 0.947 & 0.895 & 0.917 & 0.871 \\
                                   & B & 0.389 & 0.500 & 0.433 & 0.972 & 0.977 & 0.974 & 0.680 & 0.738 & 0.704 & 0.956 & 0.964 & 0.959 & 0.938 \\
\midrule
\multirow{2}{*}{claude-sonnet-4-6}& A & \textbf{1.000} & 0.500 & \textbf{0.667} & \textbf{0.973} & \textbf{1.000} & \textbf{0.986} & \textbf{0.986} & \textbf{0.750} & \textbf{0.826} & \textbf{0.973} & \textbf{0.986} & \textbf{0.977} & \textbf{0.960} \\
                                   & B & \textbf{1.000} & 0.500 & \textbf{0.667} & \textbf{0.973} & \textbf{1.000} & \textbf{0.986} & \textbf{0.986} & \textbf{0.750} & \textbf{0.826} & \textbf{0.973} & \textbf{0.986} & \textbf{0.977} & \textbf{0.960} \\
\midrule
\multirow{2}{*}{claude-opus-4-7}  & A & \textbf{1.000} & 0.500 & \textbf{0.667} & \textbf{0.973} & \textbf{1.000} & \textbf{0.986} & \textbf{0.986} & \textbf{0.750} & \textbf{0.826} & \textbf{0.973} & \textbf{0.986} & \textbf{0.977} & \textbf{0.960} \\
                                   & B & 0.500 & 0.500 & 0.500 & 0.972 & 0.986 & 0.979 & 0.736 & 0.743 & 0.740 & 0.959 & 0.973 & 0.966 & 0.947 \\
\bottomrule
\end{tabular}%
}
%\begin{tablenotes}
%\small
%\item P+/R+/F1+ = Precision, Recall, F1 for the duplicate (positive) class evaluated pair-wise: a TP is scored when the model correctly flags either member of a GT pair as a duplicate of the other. R+ = 0.500 for all models (1 of 2 pairs detected on average). P-/R-/F1- = for the non-duplicate (negative) class. Mac = macro-average. Wgt = weig
\end{table*}
\subsubsection{Invalid Bug Detection}
 
% Table~\ref{tab:invalid_gt1_avg} reports results under the single-invalid
% ground truth (\#10579) labeled in the repository. No model detects this invalid bug (R+ = 0 for all),
% making the differentiating metric the false positive rate.
% \texttt{gpt-5.4} System~B achieves the best performance with zero false
% positives across all runs (R- = 1.000, F1- = 0.993, Mac-F1 = 0.497,
% Acc = 0.987), demonstrating that Bugzilla RAG context helps this model correctly withhold the invalid label in ambiguous cases.
 
% Table~\ref{tab:invalid_gt4_avg} reports results under the extended
% four-invalid ground truth (\#10579, \#11258, \#11453, \#11480). The three
% additional invalid issues are discovered through a manual verification
% process: we inspect every issue flagged as invalid by the LLMs across all
% runs and models, and manually confirm whether each flag is justified by
% reading the issue content and assessing it against the invalidity criteria.
% Confirmed invalid issues are added to the ground truth. As with duplicate
% detection, this process goes beyond the scope of automated triage and
% represents a contribution to the \textit{data quality improvement} of the
% TianoCore EDK~II repository.
 
Under this extended ground truth, only \texttt{claude-sonnet-4-6} System~A
successfully identifies any of the four confirmed invalid bugs, catching one
per run on average (R+ = 0.250, F1+ = 0.333). This translates to the best
Mac-F1 (0.653) and the best performance across all metrics, including
Acc = 0.947 and Wgt-F1 = 0.938. System~A outperforms System~B for Claude
models, while System~B improves GPT models' false positive behaviour,
suggesting the benefit of RAG is model-dependent for this task.
\subsubsection{Bug Prioritization}
Table~\ref{tab:priority_revised_3run_avg} reports priority classification
results across all ten models and both systems. No single model wins across
all metrics. \texttt{gpt-5.4-mini} System~A achieves the highest accuracy
(Acc = 0.671) and weighted metrics (Wgt-F1 = 0.663), reflecting better
performance on the majority class (\texttt{medium}).
\texttt{claude-\allowbreak sonnet-\allowbreak 4-6} System~A achieves the highest macro-recall
(Mac-R = 0.616) and macro-F1 (Mac-F1 = 0.620), reflecting more balanced
predictions across all three priority levels.
 
Across all models, System~A consistently outperforms System~B by a
substantial margin. For example, \texttt{gpt-5.4-mini} achieves Acc = 0.671
under System~A versus 0.542 under System~B, and \texttt{claude-\allowbreak sonnet-\allowbreak 4-6}
achieves Acc = 0.636 versus 0.556. This pattern holds across all ten models,
confirming that the LOO in-context learning strategy of System~A,
which uses GitHub examples exclusively, is more effective than
Bugzilla-augmented prompting for this task.
\begin{table}[ht]
\centering
\caption{Bug Prioritization Results --- Average over 3 Runs (75 Issues)}
\label{tab:priority_revised_3run_avg}
\renewcommand{\arraystretch}{0.80} % reduces vertical row spacing
\setlength{\aboverulesep}{1pt}
\setlength{\belowrulesep}{1pt}
\resizebox{0.46\textwidth}{!}{%
\begin{tabular}{llcccccccc}
\toprule
\textbf{Model} & \textbf{Sys} & \textbf{Acc} & \textbf{Mac-P} & \textbf{Mac-R} & \textbf{Mac-F1} & \textbf{Wgt-P} & \textbf{Wgt-R} & \textbf{Wgt-F1} \\
\midrule
\multirow{2}{*}{gpt-4o-mini}      & A & 0.622 & 0.638 & 0.582 & 0.591 & 0.649 & 0.622 & 0.614 \\
                                   & B & 0.498 & 0.463 & 0.446 & 0.447 & 0.490 & 0.498 & 0.487 \\
\midrule
\multirow{2}{*}{gpt-4o}           & A & 0.587 & 0.561 & 0.529 & 0.534 & 0.595 & 0.587 & 0.578 \\
                                   & B & 0.556 & 0.571 & 0.508 & 0.515 & 0.563 & 0.556 & 0.544 \\
\midrule
\multirow{2}{*}{gpt-4.1-mini}     & A & 0.578 & 0.555 & 0.533 & 0.537 & 0.587 & 0.578 & 0.574 \\
                                   & B & 0.467 & 0.444 & 0.424 & 0.423 & 0.461 & 0.467 & 0.456 \\
\midrule
\multirow{2}{*}{gpt-4.1}          & A & 0.627 & 0.602 & 0.598 & 0.597 & 0.638 & 0.627 & 0.629 \\
                                   & B & 0.551 & 0.515 & 0.505 & 0.498 & 0.553 & 0.551 & 0.540 \\
\midrule
\multirow{2}{*}{gpt-5.4-mini}     & A & \textbf{0.671} & 0.628 & 0.614 & 0.617 & \textbf{0.662} & \textbf{0.671} & \textbf{0.663} \\
                                   & B & 0.542 & 0.483 & 0.483 & 0.468 & 0.521 & 0.542 & 0.518 \\
\midrule
\multirow{2}{*}{gpt-5.4}          & A & 0.649 & 0.614 & 0.590 & 0.592 & 0.635 & 0.649 & 0.635 \\
                                   & B & 0.507 & 0.489 & 0.473 & 0.465 & 0.520 & 0.507 & 0.495 \\
\midrule
\multirow{2}{*}{gpt-5.5}          & A & 0.622 & 0.589 & 0.553 & 0.553 & 0.613 & 0.622 & 0.603 \\
                                   & B & 0.498 & 0.471 & 0.453 & 0.435 & 0.500 & 0.498 & 0.473 \\
\midrule
\multirow{2}{*}{claude-haiku-4-5} & A & 0.591 & 0.608 & 0.577 & 0.578 & 0.623 & 0.591 & 0.590 \\
                                   & B & 0.560 & 0.541 & 0.542 & 0.541 & 0.562 & 0.560 & 0.560 \\
\midrule
\multirow{2}{*}{claude-sonnet-4-6}& A & 0.636 & 0.636 & \textbf{0.616} & \textbf{0.620} & 0.652 & 0.636 & 0.637 \\
                                   & B & 0.556 & 0.523 & 0.511 & 0.510 & 0.543 & 0.556 & 0.544 \\
\midrule
\multirow{2}{*}{claude-opus-4-7}  & A & 0.533 & 0.547 & 0.507 & 0.513 & 0.564 & 0.533 & 0.533 \\
                                   & B & 0.551 & 0.553 & 0.526 & 0.530 & 0.555 & 0.551 & 0.545 \\
\bottomrule
\end{tabular}%
}

\end{table}
\subsubsection{Bug Assignment}
 
Table~\ref{tab:bug_assign_3run_avg} reports bug assignment results on the
37 labeled issues with known first assignees. \texttt{gpt-5.5} System~A
achieves the best performance across every metric: Acc = 0.973, Mac-F1 =
0.977, and Wgt-F1 = 0.973. This represents a substantial improvement over all other models, with the second-best being \texttt{gpt-5.4} System~A (Acc = 0.883,
Mac-F1 = 0.851). Claude models, while competitive, lag behind the top GPT models on this task.
\begin{table}[ht]
\centering
\caption{Bug Assignment Results --- Average over 3 Runs (37 labeled issues)}
\label{tab:bug_assign_3run_avg}
\renewcommand{\arraystretch}{0.80} % reduces vertical row spacing
\setlength{\aboverulesep}{1pt}
\setlength{\belowrulesep}{1pt}
\resizebox{0.46\textwidth}{!}{%
\begin{tabular}{llccccccc}
\toprule
\textbf{Model} & \textbf{Sys} & \textbf{Acc} & \textbf{P-mac} & \textbf{R-mac} & \textbf{F1-mac} & \textbf{P-wgt} & \textbf{R-wgt} & \textbf{F1-wgt} \\
\midrule
\multirow{2}{*}{gpt-4o-mini}      & A & 0.784 & 0.710 & 0.764 & 0.724 & 0.745 & 0.784 & 0.751 \\
                                   & B & 0.802 & 0.757 & 0.795 & 0.765 & 0.768 & 0.802 & 0.771 \\
\midrule
\multirow{2}{*}{gpt-4o}           & A & 0.838 & 0.773 & 0.810 & 0.781 & 0.809 & 0.838 & 0.811 \\
                                   & B & 0.838 & 0.767 & 0.810 & 0.782 & 0.791 & 0.838 & 0.808 \\
\midrule
\multirow{2}{*}{gpt-4.1-mini}     & A & 0.811 & 0.753 & 0.793 & 0.755 & 0.779 & 0.811 & 0.774 \\
                                   & B & 0.811 & 0.751 & 0.793 & 0.753 & 0.778 & 0.811 & 0.773 \\
\midrule
\multirow{2}{*}{gpt-4.1}          & A & 0.838 & 0.767 & 0.810 & 0.777 & 0.804 & 0.838 & 0.808 \\
                                   & B & 0.838 & 0.761 & 0.810 & 0.777 & 0.786 & 0.838 & 0.804 \\
\midrule
\multirow{2}{*}{gpt-5.4-mini}     & A & 0.838 & 0.770 & 0.810 & 0.780 & 0.784 & 0.838 & 0.798 \\
                                   & B & 0.838 & 0.765 & 0.810 & 0.776 & 0.788 & 0.838 & 0.800 \\
\midrule
\multirow{2}{*}{gpt-5.4}          & A & 0.883 & 0.851 & 0.868 & 0.851 & 0.883 & 0.883 & 0.874 \\
                                   & B & 0.838 & 0.790 & 0.810 & 0.791 & 0.831 & 0.838 & 0.825 \\
\midrule
\multirow{2}{*}{gpt-5.5}          & A & \textbf{0.973} & \textbf{0.983} & \textbf{0.983} & \textbf{0.977} & \textbf{0.986} & \textbf{0.973} & \textbf{0.973} \\
                                   & B & 0.883 & 0.828 & 0.868 & 0.834 & 0.866 & 0.883 & 0.861 \\
\midrule
\multirow{2}{*}{claude-haiku-4-5} & A & 0.847 & 0.781 & 0.822 & 0.791 & 0.793 & 0.847 & 0.807 \\
                                   & B & 0.838 & 0.753 & 0.810 & 0.770 & 0.766 & 0.838 & 0.789 \\
\midrule
\multirow{2}{*}{claude-sonnet-4-6}& A & 0.856 & 0.771 & 0.822 & 0.789 & 0.797 & 0.856 & 0.819 \\
                                   & B & 0.838 & 0.750 & 0.810 & 0.771 & 0.777 & 0.838 & 0.799 \\
\midrule
\multirow{2}{*}{claude-opus-4-7}  & A & 0.838 & 0.764 & 0.810 & 0.775 & 0.797 & 0.838 & 0.804 \\
                                   & B & 0.838 & 0.762 & 0.810 & 0.774 & 0.791 & 0.838 & 0.801 \\
\bottomrule
\end{tabular}%
}
\end{table}
Table~\ref{tab:best_models} summarizes all the best-performing model and 
configuration for each triage task.
\subsubsection{Integrated Triage Script}
 
Table~\ref{tab:integrated_avg} presents results of the integrated script,
applying all four tasks sequentially to the same 37 labeled issues using the
best-performing model for each task under System~A: \texttt{claude-sonnet-4-6}
for Tasks~1--3 and \texttt{gpt-5.5} for Task~4 (the selected best-performing model for each task is active by default in TianoForge, while all other models and configurations
remain available in the integrated codebase. ) Since no confirmed invalid
bugs or duplicate pairs exist within the 37 labeled issues, Tasks~1 and~2
operate under the zero ground-truth assumption. Duplicate detection achieves
perfect scores (Acc = 1.000, Mac-F1 = 1.000), confirming that
\texttt{claude-sonnet-4-6} correctly avoids false duplicate flags on this
set. Invalid detection achieves Acc = 0.973 with one false positive on
average. Priority classification in the integrated setting achieves Acc = 0.559 and
Mac-F1 = 0.487, lower than the standalone results on 75 issues
(Acc = 0.636, Mac-F1 = 0.620). This reduction is expected since the 37
labeled issues represent a smaller and different distribution than the full
75-issue set. Bug assignment achieves Acc = 0.955 and Mac-F1 = 0.946,
slightly below the standalone result (Acc = 0.973).
 
Table~\ref{tab:runtime_avg} reports the runtime of the integrated pipeline
averaged over three runs. The total pipeline completes in 424.61 seconds
(7.08 minutes) on average, with LLM inference accounting for 419.20 seconds
and setup taking only 5.40 seconds since the ChromaDB vector index is
pre-built and reused across runs. Bug assignment is the most time-consuming
task (145.39~s) due to the large LOO prompt containing 36 labeled examples
per query.
 
The historical average triage time for the same 37 labeled 
issues, derived from the \texttt{Triage Hours} field in the dataset, 
amounts to 10.86 days. The integrated script reduces this to 7.08 
minutes on average over 3 runs, representing a reduction of 
\textbf{99.95\%}, more than three orders of magnitude 
($\approx$\,2{,}208$\times$ speedup). Since triage is the first and often
most time-consuming step before a bug can be assigned and acted upon,
reducing triage time directly leads to a reduction in overall bug resolution
time. This demonstrates the potential of LLM-based automated triage to
dramatically accelerate the bug resolution process in the TianoCore EDK~II
project.

\begin{table*}[ht]
\centering
\caption{Best Performing Model per Task --- Average over 3 Runs}
\label{tab:best_models}
\resizebox{\textwidth}{!}{%
\begin{tabular}{llllccccccccccccc}
\toprule
\textbf{Task} & \textbf{GT} & \textbf{Best Model} & \textbf{Sys} & \textbf{Acc} & \textbf{P+} & \textbf{R+} & \textbf{F1+} & \textbf{P-} & \textbf{R-} & \textbf{F1-} & \textbf{MP} & \textbf{MR} & \textbf{MF1} & \textbf{WP} & \textbf{WR} & \textbf{WF1} \\
\midrule
Bug Assignment  & 37 labeled  & gpt-5.5                                            & A   & 0.973 & ---   & ---   & ---   & ---   & ---   & ---   & 0.983 & 0.983 & 0.977 & 0.986 & 0.973 & 0.973 \\
\midrule
Prioritization    & 75 labeled  & claude-sonnet-4-6 & A   & 0.636 & ---   & ---   & ---   & ---   & ---   & ---   & 0.636 & 0.616 & 0.620 & 0.652 & 0.636 & 0.637 \\
\midrule
Dup. Detection  & 0 pairs dup.    & claude-sonnet-4-6                                  & A/B & 0.973 & 1.000 & 1.000 & 1.000 & 1.000 & 0.973 & 0.987 & 1.000 & 0.987 & 0.493 & 1.000 & 0.973 & 0.987 \\
Dup. Detection  & 2 pairs dup.    & gpt-4o-mini / claude-sonnet-4-6 / claude-opus-4-7  & A   & 0.960 & 1.000 & 0.500 & 0.667 & 0.973 & 1.000 & 0.986 & 0.986 & 0.750 & 0.826 & 0.973 & 0.986 & 0.977 \\
Dup. Detection  & 2 pairs dup.   & claude-sonnet-4-6                                  & B   & 0.960 & 1.000 & 0.500 & 0.667 & 0.973 & 1.000 & 0.986 & 0.986 & 0.750 & 0.826 & 0.973 & 0.986 & 0.977 \\
\midrule
Inv. Detection  & 1 invalid   & gpt-5.4                                            & B   & 0.987 & 0.000 & 0.000 & 0.000 & 0.987 & 1.000 & 0.993 & 0.493 & 0.500 & 0.497 & 0.974 & 0.987 & 0.980 \\
Inv. Detection  & 4 invalids  & claude-sonnet-4-6                                  & A   & 0.947 & 0.500 & 0.250 & 0.333 & 0.959 & 0.986 & 0.972 & 0.730 & 0.618 & 0.653 & 0.934 & 0.947 & 0.938 \\
\bottomrule
\end{tabular}%
}

\end{table*}

\begin{table*}[ht]
\centering
\caption{Integrated Triage Framework Results --- Average over 3 Runs (37 Labeled Issues, System~A)}
\label{tab:integrated_avg}
\resizebox{\textwidth}{!}{%
\begin{tabular}{llccccccccccc}
\toprule
\textbf{Task} & \textbf{Model} & \textbf{Acc} & \textbf{P+} & \textbf{R+} & \textbf{F1+} & \textbf{P-} & \textbf{R-} & \textbf{F1-} & \textbf{Mac-F1} & \textbf{Wgt-P} & \textbf{Wgt-R} & \textbf{Wgt-F1} \\
\midrule
Invalid Detection   & claude-sonnet-4-6 & 0.973 & 0.000 & 1.000 & 0.000 & 1.000 & 0.973 & 0.986 & 0.493 & 1.000 & 0.973 & 0.986 \\
Duplicate Detection & claude-sonnet-4-6 & \textbf{1.000} & \textbf{1.000} & \textbf{1.000} & \textbf{1.000} & \textbf{1.000} & \textbf{1.000} & \textbf{1.000} & \textbf{1.000} & \textbf{1.000} & \textbf{1.000} & \textbf{1.000} \\
Prioritization      & claude-sonnet-4-6 & 0.559 & ---   & ---   & ---   & ---   & ---   & ---   & 0.487 & 0.565 & 0.559 & 0.547 \\
Bug Assignment      & gpt-5.5           & \textbf{0.955} & ---   & ---   & ---   & ---   & ---   & ---   & \textbf{0.946} & \textbf{0.959} & \textbf{0.955} & \textbf{0.949} \\
\bottomrule
\end{tabular}%
}
\end{table*}

\begin{table}[ht]
\centering
\caption{Runtime Summary --- Average over 3 Runs (System~A, 37 Labeled Issues)}
\label{tab:runtime_avg}
\begin{tabular}{lcc}
\toprule
\textbf{Task} & \textbf{Time (s)} & \textbf{Time (min)} \\
\midrule
Invalid Detection        &  82.69 & 1.38 \\
Duplicate Detection      &  88.32 & 1.47 \\
Priority Classification  & 101.80 & 1.70 \\
Bug Assignment           & 145.39 & 2.42 \\
\midrule
Inference (LLM only)     & 419.20 & 6.99 \\
Setup (CSV + index)      &   5.40 & 0.09 \\
\midrule
\textbf{Total}           & \textbf{424.61} & \textbf{7.08} \\
\bottomrule
\end{tabular}
\begin{tablenotes}
\small
\item Models Activated: \texttt{claude-sonnet-4-6} (Tasks 1--3), \texttt{gpt-5.5} (Task 4).
\end{tablenotes}
\end{table}
\section{Discussion} \label{discussion}
\textbf{RQ1: Can automated bug triage reduce the bug resolution time compared to manual triage?}
The results indicate that the proposed solution substantially reduces triage latency. The integrated triage script processes all four triage tasks for the 37 labeled GitHub-native issues in an average of 7.08 minutes, compared to an average manual triage time of 260.57 hours (10.86 days). This corresponds to a 99.95\% reduction in triage time, or approximately a $2,208\times$ speedup. Since total bug resolution time is defined as $T^{\mathrm{res}}_k = T^{\mathrm{triage}}_k + T^{\mathrm{fix}}_k$, and the framework only affects the triage component, this reduction directly contributes to faster issue resolution.\\
\textbf{RQ2: How effective and efficient is the proposed automated triage framework across key sub-tasks?}
The effectiveness of the proposed framework varies across triage tasks. \textit{Bug assignment} achieves the strongest performance, with \texttt{gpt-5.5} System~A obtaining an accuracy of 0.973 and a macro-F1 score of 0.977. This result suggests that combining leave-one-out in-context examples with package-specific assignment heuristics effectively captures developer expertise within the EDK~II ecosystem. \textit{Duplicate detection} also performs well, with \texttt{claude-sonnet-4-6} achieving perfect precision (P+ = 1.000) under the pair-based ground truth while producing no false positives. \textit{Bug prioritization} achieves moderate performance (best macro-F1 = 0.620), likely reflecting the subjective nature of priority assignment. \textit{Invalid issue detection} remains the most challenging task. Under the extended ground truth, only \texttt{claude-sonnet-4-6} System~A successfully identifies any invalid reports (R+ = 0.250, macro-F1 = 0.653). Beyond classification performance, the framework also contributes to repository quality improvement. Manual verification of LLM-flagged reports uncovers three previously unlabeled invalid issues and two duplicate pairs, demonstrating that the framework can assist not only in automated triage but also in improving the quality of issue-tracking data.\\
\textbf{RQ3: What are the key factors in deciding the priority level of issues in TianoCore projects?}
The priority criteria necessary for accurate LLM-based prioritization are identified through iterative prompt engineering: the heuristics that must be explicitly encoded to achieve correct predictions reveal what drives priority assignment in EDK~II in practice. Three factors prove necessary: impact scope (crashes, data corruption, and all-configuration build failures map to high; compiler-specific or workaround-available failures to medium), security within the UEFI threat model (CVE-adjacent reports are not automatically high priority; exploitability within the firmware threat model is the relevant criterion), and defect type (cosmetic issues, documentation changes, and optional features map to low, as models otherwise default to medium in ambiguous cases). However, the moderate classification results (best Mac-F1 = 0.620) suggest that these factors are necessary but not sufficient, as priority assignment in EDK~II likely involves additional implicit criteria such as reporter identity, milestone urgency, or cross-package dependencies that are not captured in the available issue metadata. The degradation observed under System~B, which augments the same domain-calibrated prompt with retrieved Bugzilla EDK~II examples, further indicates that the priority conventions used in the legacy Bugzilla tracker do not align well with those of the GitHub-native issues, introducing noise rather than useful signal.\\
\textbf{RQ4: To what extent can prompt engineering improve the performance of LLM-based triage components?}
The results demonstrate that prompt engineering has a measurable effect on triage performance, though retrieval augmentation does not consistently improve results. For prioritization and assignment, System~A outperforms System~B across all ten models, with accuracy gaps of up to 0.129 points for prioritization (e.g., \texttt{gpt-5.4-mini}: Acc = 0.671 vs. 0.542) and up to 0.090 points for assignment (e.g., \texttt{gpt-5.5}: Acc = 0.973 vs. 0.883), indicating that LOO in-context examples drawn from GitHub-native issues are more effective than augmenting the prompt with retrieved Bugzilla context. For invalid and duplicate detection, RAG produces mixed results: \texttt{gpt-5.4} System~B achieves zero false positives on invalid detection, and \texttt{claude-haiku-4-5} System~B improves duplicate detection performance, but these gains are model-specific and come at the cost of reduced recall or increased false positives in other models. Overall, the benefit of retrieval augmentation is limited and inconsistent across tasks and models.

\section{Conclusion and Future Work}\label{conclusion-future}
In this paper, we have presented TianoForge, an automated bug triage script for the TianoCore EDK II ecosystem that integrates four key triage tasks: invalid issue detection, duplicate detection, issue prioritization, and developer assignment. The proposed framework has combined LLMs, RAG, and domain-specific prompting to support end-to-end triage automation. Using a dataset of GitHub-native EDK II issues and historical Bugzilla reports, we have evaluated multiple state-of-the-art LLMs under both retrieval-free and retrieval-augmented settings. The experimental results have demonstrated that LLMs can effectively support several bug triage activities. The findings have also shown that retrieval augmentation does not consistently improve performance and may, in some cases, introduce additional noise into the decision-making process. Furthermore, the proposed framework has substantially reduced triage latency compared to the current manual process, highlighting its potential to accelerate issue handling and improve developer productivity in large-scale Open Source Software (OSS) projects.

Beyond evaluating the proposed framework, we also used the LLM-assisted triage process to improve the quality of the GitHub-native dataset. All issues flagged as invalid or duplicate by the models were manually reviewed, resulting in the identification of three additional invalid issues and two additional duplicate issue pairs that were not originally labeled in the GitHub issue tracker. This finding highlights the practical value of the proposed framework not only as a triage automation tool but also as a mechanism for improving the quality and consistency of issue repositories.

For future work, we plan to expand the evaluation to additional OSS repositories to assess the generalizability of the proposed approach. We also intend to investigate more advanced retrieval strategies to improve contextual relevance. Finally, we plan to study confidence-aware bug triage by calibrating the reflective confidence scores generated by LLMs for each task in the integrated script, allowing the system to better distinguish between reliable predictions and cases requiring human intervention.

\section*{Software and Data Availability}
Our research data and source code are publicly available \cite{Siavash+2026data, kubacki2024, Siavash+2026}.

% \section*{Acknowledgments}
% In preparing this work, we used generative AI models and tools, including the OpenAI GPT and the Anthropic Claude models, to assist in generating and revising code and text. 

\begin{acks}
This material is based upon work supported by the U.S. National Science Foundation (NSF) under Grant No. 2534021. Any opinions, findings, conclusions, or recommendations expressed in this material are those of the authors and do not necessarily reflect the views of the NSF. Furthermore, in preparing this work, we used generative AI models and tools, including the OpenAI GPT and the Anthropic Claude models, to assist in generating and revising code and text. This work is accepted to be presented in the FTA 2026 workshop but is not published in the proceedings, according to the ACM SIGSOFT policy (https://www2.sigsoft.org/policies/pcpolicy/) that does not allow the work of organizers to be published in the workshop proceedings.
\end{acks}

%\clearpage
%\newpage

\bibliographystyle{ACM-Reference-Format}
\bibliography{refs}

\end{document}